\documentclass[aps,prb,twocolumn,showpacs,groupedaddress]{revtex4}

\def\comment#1{\bigskip\hrule\smallskip#1\smallskip\hrule\bigskip}   
\usepackage{amsmath}
\usepackage{graphicx}
\usepackage{dcolumn}
\usepackage{bm}

\begin{document}


\title{Quantum Monte Carlo simulation of spin-polarized H}

\author{L. Vranje\v{s} Marki{\'c}}
\affiliation{Faculty of Natural Sciences, University of Split, 21000 Split,
Croatia\\
Institut f\"ur Theoretische Physik, 
Johannes Kepler Universit\"at, A 4040 Linz, Austria}
\author{J. Boronat}
\author{J. Casulleras}
\affiliation{Departament de F\'\i sica i Enginyeria Nuclear, Campus Nord
B4-B5, Universitat Polit\`ecnica de Catalunya, E-08034 Barcelona, Spain}
\def\comment#1{\bigskip\hrule\smallskip#1\smallskip\hrule\bigskip}   
\date{\today}

\begin{abstract}
The ground-state properties of spin polarized hydrogen H$\downarrow$ are
obtained  by means of diffusion Monte Carlo calculations. Using the most 
accurate to date ab
initio H$\downarrow$-H$\downarrow$ interatomic potential we have studied
its gas phase, from the very dilute regime until densities above 
its freezing point. At very
small densities, the equation of state of the gas is very well described in
terms of the gas parameter $\rho a^3$, with $a$ the s-wave scattering
length. The solid phase has also been studied up to high pressures. The
gas-solid phase transition occurs at a pressure of 173 bar, a much
higher value than suggested by previous approximate descriptions.

\end{abstract}

\pacs{67.65.+z,02.70.Ss}

\maketitle

\section{Introduction}\

The suggestion of Stwalley and Nosanow~\cite{StwaleyNosanow} in 1976 that
electron-spin-polarized gases of hydrogen (H$\downarrow$) could be an
ideal candidate for achieving a Bose-Einstein condensate (BEC) state opened
an intense experimental search which  finally led to its first observation
in Rb, Na and Li in
1995.~\cite{becalkali} The success attained in cold alkali gases was made possible due to
techniques like evaporative cooling developed previously for confining   
H$\downarrow$. The realization of a BEC state of hydrogen was finally
achieved in 1998 by Fried \textit{et al.}~\cite{fried} after overcoming
arduous problems like recombination on the walls, by working with a
wall-free confinement, and low evaporation rates by using spin
resonance. Also in 1998, Safonov \textit{et al.}~\cite{safonov} observed a
quasi-condensate in two-dimensional (2D) H$\downarrow$ adsorbed on liquid
$^4$He. This is one of the best realizations of a 2D quantum system since
the adsorption energy of H$\downarrow$ on $^4$He is only $\sim 1$ K and the
adsorbed gas floats approximately 8 \AA \ apart from the liquid. Further
experimental work is still necessary to observe in this system signals of
the Berezinskii-Kosterlitz-Thouless transition, which has been
recently observed in a trapped gas of Rb confined in such a way that atoms
can move only within a plane.~\cite{dalibard} 

Hydrogen is the simplest element and its main properties are well known
theoretically, starting from the interatomic interaction which can
be computed almost exactly.~\cite{kolos} This is significantly different from 
alkali gases in which the interaction is much more involved and, in
general, less well known. Its s-wave scattering length $a$ is appreciably
smaller than the typical values for alkalis, a feature that retards
evaporative cooling and produces a higher transition temperature 
(50 $\mu K$). Spin-polarized hydrogen atoms interact via the triplet
potential $b$ $^3\Sigma_u^+$ determined in an essentially exact way by Kolos
and Wolniewicz,~\cite{kolos} and recently extended to larger interparticle
distances by Jamieson \textit{et al.}.~\cite{jamieson} The
H$\downarrow$-H$\downarrow$ interaction is highly repulsive at short
distances and presents a shallow minimum of $\sim 6$K at $r \sim 4$\AA.
The combination of this extremely weak attraction and its light mass
explains why  H$\downarrow$ remains in the gas phase even in the limit of zero
temperature. A measure of the quantum character of a given system can be
drawn through the quantum parameter,~\cite{miller}
\begin{equation}
\eta = \frac{\hbar^2}{m \epsilon \sigma^2} \ ,
\label{deboer}
\end{equation}
with $\epsilon$ and $\sigma$ the well depth and core of the interaction,
respectively. According to this definition, $\eta=0.5$ for H$\downarrow$
which is the highest value for $\eta$ among all the quantum fluids 
(for instance, $\eta=0.2$ for $^4$He).

From a theoretical viewpoint, bulk H$\downarrow$ was first studied by
Etters \textit{et al.}~\cite{etters} using the variational Monte Carlo (VMC) 
method and by
Miller and Nosanow~\cite{miller} using integral equations for computing the
multidimensional integrals of the variational approach. Both calculations
showed that the ground state of H$\downarrow$ is unbound at any pressure.
Lantto and Nieminen~\cite{lantto} confirmed this result using the 
Euler-Lagrange hypernetted
chain equation and estimated values for the condensate
fraction of the gas at different densities for the first time. More
recently, Entel and Anlauf~\cite{entel} carried out a new VMC calculation of 
properties
of the gas phase such as the energy, condensate fraction and excitation
spectrum. The heavier isotopes spin-polarized deuterium D$\downarrow$ and
tritium T$\downarrow$ have also been studied theoretically. D$\downarrow$
atoms obey Fermi statistics and their three versions, involving one
(D$\downarrow_1$), two (D$\downarrow_2$)  and
three  (D$\downarrow_3$)  equally occupied nuclear spin states, were  
analyzed by Panoff and 
Clark~\cite{panoff} and Flynn \textit{et al.}~\cite{flynn} 
using both VMC and Fermi-hypernetted chain theory
(FHNC). From the results obtained, they concluded that D$\downarrow_2$ and
D$\downarrow_3$ are both liquids at zero pressure. Their conclusion has been confirmed by Skjetne and \O stgaard~\cite{skjetne}, using a lowest-order constrained variational method. On the other hand,
microscopic properties of bosonic tritium  T$\downarrow$ clusters have been
recently studied by Blume \textit{et al.}~\cite{blume} using the diffusion
Monte Carlo (DMC) method, and their results suggest the use of  
T$\downarrow$ as a new BEC gas with the advantage of a nearly exact 
knowledge of the interatomic potential.

In the present work, we present a DMC study of the gas and solid phases of
spin-polarized hydrogen. Using recent updates of the \textit{ab initio}
H$\downarrow$-H$\downarrow$ interatomic potential and relying on the  
accuracy of the
DMC method, we report accurate microscopic results for energetic and structural
properties of the bulk system. In the very low density regime, the energy
is well reproduced by the well-know analytical expansion in terms of the
gas parameter $\rho a^3$,~\cite{universal1,universal2,universal3} 
with $a$ the s-wave scattering length 
obtained by
solving the two-body Schr\"odinger equation with the chosen interatomic
potential. A relevant result of our work is an accurate computation of the
gas-solid phase transition point which is predicted to occur at pressures
significantly higher than previous predictions based on quantum theory of
corresponding states~\cite{StwaleyNosanow} and VMC simulations.~\cite{danilowicz}   

In Sec. II, we report briefly the DMC method and discuss the trial wave
functions used for importance sampling in the gas and solid phases. In Sec.
III, the results of the DMC simulations are reported in several
subsections. In the first one, we review the H$\downarrow$-H$\downarrow$ 
interatomic potentials and
compare our results at the variational level with some previous
estimations. The second and third parts are devoted to the microscopic
results for the gas and solid phases, respectively. In the last one, we
study the gas-solid phase transition point and report results on the
freezing and melting densities. Finally, Sec. IV comprises a summary of the
work and an account of the main conclusions.

\section{Method}

The DMC method is nowadays a well-known  tool devised to study quantum fluids 
and
solids at zero temperature. Its starting point is the Schr\"odinger equation 
written
in imaginary time,
\begin{equation}
-\hbar \frac{\partial \Psi(\bm{R},t)}{\partial t} = (H- E_{\text r}) \Psi(\bm{R},t) \ ,
\label{srodin}
\end{equation}
with an $N$-particle Hamiltonian
\begin{equation}
 H = -\frac{\hbar^2}{2m} \sum_{i=1}^{N} \bm{\nabla}_i^2 + \sum_{i<j}^{N} V(r_{ij})  \ .
\label{hamilto}
\end{equation}
In Eq. (\ref{srodin}), $E_{\text r}$ is a constant acting as a reference energy and
$\bm{R} \equiv (\bm{r}_1,\ldots,\bm{r}_N)$ is a \textit{walker} in Monte
Carlo therminology. 

DMC solves stochastically the Schr\"odinger
equation (\ref{srodin}) replacing $\Psi(\bm{R},t)$ by $\Phi(\bm{R},t)=
\Psi(\bm{R},t) \psi(\bm{R})$,
with $\psi(\bm{R})$ a trial wave function used for importance sampling. 
In this way equation (\ref{srodin}) becomes
\begin{eqnarray}
-\frac{\partial \Phi(\bm{R},t)}{\partial t}  & = &  -D\,
\bm{\nabla}^2_{{\bf R}}
\Phi(\bm{R},t)+D\, \bm{\nabla}_{\bm{R}} \left( \bm{F}(\bm{R})
\,\Phi(\bm{R},t)\,
\right) \nonumber \\ 
& & 
+\left(E_L(\bm{R})-E \right)\,\Phi(\bm{R},t)  \ ,
  \label{dmceq}
\end{eqnarray}
where $D=\hbar^2 /(2m)$, $E_L(\bm{R})=\psi(\bm{R})^{-1} H \psi(\bm{R})$
is the local energy, and
\begin{equation}
\bm{F}(\bm{R}) = 2\, \psi(\bm{R})^{-1}
\bm {\nabla}_{\bm{R}} \psi(\bm{R})
\label{dmceq2}
\end{equation}
is the drift force which guides the diffusion process.
In Eq. (\ref{dmceq}), when $t \rightarrow \infty$ only the lowest energy 
eigenfunction, not
orthogonal to $\psi(\bm{R})$, survives and then the sampling of the ground
state is effectively achieved. Apart from statistical uncertainties, the
energy of a $N$-body bosonic system is exactly calculated.

The trial wave function used for the simulation of the gas phase is of
Jastrow form, $\psi_{\text J}(\bm{R}) = \prod_{i<j}^{N} f(r_{ij})$, with a
two-body correlation function $f(r)$ of the form
\begin{equation}
f(r)=\exp [-b_1 \exp(-b_2r )] \ ,
\label{trial}
\end{equation}
where $b_1$ and $b_2$ are variational parameters. This form has been taken
from the VMC work of Etters \textit{et al.},~\cite{etters} who used a Morse
potential fitted to reproduce Kolos and Wolniewicz \textit{ab
initio} data.~\cite{kolos} It corresponds
to the WKB solution of the two-body Schr\"odinger equation for small
interparticle distances when the potential is of Morse type.    

Simulations of the crystalline
bcc, fcc and hcp phases have been also carried out; in this case, we use a Nosanow-Jastrow 
model
\begin{equation}
\psi_{\text{NJ}}(\bm{R}) = \psi_{\text J}(\bm{R}) 
\prod_{i}^{N}g(r_{iI}) \ ,
\label{nosanow}
\end{equation}
 where $g(r)=\exp(-\alpha r^2/2)$ is a localizing function which links every
particle $i$ to a point $\bm{r}_I$ of the lattice. The parameter $\alpha$ is
optimized variationally. 
  
The variational parameters $b_1$, $b_2$, and $\alpha$
(\ref{trial},\ref{nosanow}) have been obtained at different densities by
optimizing the variational energy calculated with the VMC method.
For example, in the gas phase and at a density
$\rho=0.0079$ \AA$^{-3}$ the values are  $b_1=82$ and $b_2=1.32$ \AA$^{-1}$;
$b_1$ increases with density whereas $b_2$ remains practically
constant. In the solid phase the most relevant parameter is $\alpha$,  which
increases from a value 0.3 \AA$^{-2}$ at melting density up to 1.3
\AA$^{-2}$ at the highest density here studied; the Jastrow parameters are
kept fixed in all the solid density range, $b_1=70$ and
$b_2=1.32$ \AA$^{-1}$. The statistical errors of the 
variational energies are similar to those of the DMC results (see
Tables \ref{tab:Gas} and \ref{tab:BCCHydrogen}).
  
We use the DMC method accurate to second order in the time step $\Delta
t$,~\cite{boro} and then larger $\Delta t$ values than in linear DMC
algorithms can be used. We have
studied both the time-step and the mean walker population in order to eliminate
any bias coming from them. Finally, we have analyzed carefully 
the size dependence of our simulations. The calculations on the gas phase
have been carried out with 128 atoms and some checks with 150 and 170 atoms
have also been made. Using standard tail corrections, which assume a
uniform system ($g(r)=1$) beyond $r > L/2$, with $L$ the length of the
simulation box,  the size dependence of the energy remains smaller than the
typical size of the statistical errors. The size effects are larger in the
calculations of the solid phase. In this case, we have used 128, 108, and
180 atoms for the bcc, fcc, and hcp lattices. To consider tail corrections
in the same form as in the gas phase is a rough approximation 
due to the periodic order of the solid. In order to overcome this
difficulty, we have studied the size dependence of the energy at the VMC
level where larger number of particles can be used. From the VMC results one extracts
the tail corrections for a given number of atoms and then these are added to
the DMC energies. It has been verified~\cite{overpressure} that this procedure is able to
reproduce accurately the experimental equation of state of solid $^4$He.

\section{Results}

\subsection{Interatomic potential}

Spin-polarized hydrogen atoms interact via the triplet potential
$b~^3\Sigma_u^+$, calculated with high precision by Kolos and Wolniewicz
(KW) in 1965.~\cite{kolos} Due to the simplicity of the H atom 
it is possible to calculate this potential in an essentially exact way.
More recently,
it has been recalculated up to larger interatomic distances by
Jamieson, Dalgarno and Wolniewicz (JDW).~\cite{jamieson}  The differences
between the KW and JDW potentials, in the range where they can be
compared,  are rather small, as shown in Fig. 1. The addition of
mass-dependent adiabatic corrections, which have been calculated by Kolos
and Rychlewski,~\cite{kolos2} to JDW potential, can not be discerned in Fig.
1.

\begin{figure}
\centering
        \includegraphics[width=0.8\linewidth]{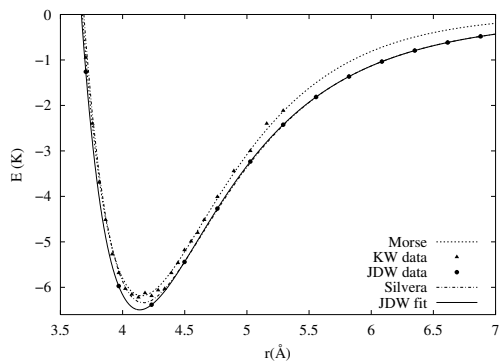}%
        \caption{  H$\downarrow$-H$\downarrow$ interatomic potentials.} 
	\label{fig:potential}
\end{figure}

In the past, only the KW potential has been used in the study of the
H$\downarrow$ gas. Usually, an analytic form was assumed and then the free
parameters of the model were fitted to reproduce the KW data. In this way,
Etters \textit{et al.}~\cite{etters} used a Morse potential whereas Silvera
and Goldman~\cite{silvera} proposed a form which is similar to the ones
used for He-He potentials. The results of these models are also plotted in
Fig. \ref{fig:potential}. In the present work, we have used the JDW interaction and a cubic
spline to interpolate between the reported points. The resulting potential
is plotted as a solid line in Fig. \ref{fig:potential}. The JDW data are smoothly connected
with the long-range behavior of the H-H potential as calculated by Yan
\textit{et al.}~\cite{yan} The JDW potential has a core diameter of 3.67
\AA\, and a minimum $\epsilon=-6.49$ K (slightly deeper than KW) 
at a distance $r_{\text m}=4.14$ \AA.      

\begin{figure}
\centering
        \includegraphics[width=0.8\linewidth]{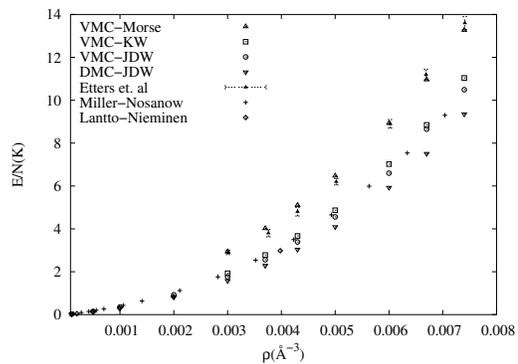}%
        \caption{Influence of the interatomic potential on the equation of
	state of gas H$\downarrow$.} 
	\label{fig:comparpot}
\end{figure}

The influence of the potential on the energy of the gas at small
densities is shown in Fig. \ref{fig:comparpot}. We have carried out VMC calculations using
the trial wave function (\ref{trial}) introduced in the previous Section
and the KW and JDW potentials. The JDW energies are below the KW ones in all
the density range, reflecting the slightly deeper well of the JDW potential.
If we use the Morse potential adjusted by Etters \textit{et
al.}~\cite{etters} to the KW
data, the VMC energies are significantly worse with respect to the KW ones.
This manifests the difficulties on fitting a functional form to the
\textit{ab initio} KW data; the Morse potential is a bit more
repulsive than KW and therefore the energies are higher. As a matter of
comparison, we also show in Fig. \ref{fig:comparpot} results from previous calculations. The
results from Etters \textit{et al.}~\cite{etters} using the Morse potential are in nice
agreement with our VMC results using the same potential. The variational
results of Miller and Nosanow~\cite{miller} used the KW data and are in close agreement
with our present VMC results with the same interaction. Finally, results of
Lantto and Nieminen~\cite{lantto} are also reported; they used the KW potential and
performed an Euler-Lagrange-HNC calculation. Their results, restricted to
very low densities, are slightly better than ours due to their use of an
optimized Jastrow factor. All these variational results, both ours and from
previous works, are compared in the same Figure with present DMC results
with the JDW interaction. As expected, the DMC results are below the VMC
ones in all the density range with a difference that increases with $\rho$,
a predictable feature attending to the fact that the Jastrow factor
(\ref{trial}) corresponds to an analytical form which approximates the wave
function solution of the two-body problem.

\subsection{Gas phase}

\begin{figure}
\centering
        \includegraphics[width=0.5\linewidth,angle=-90]{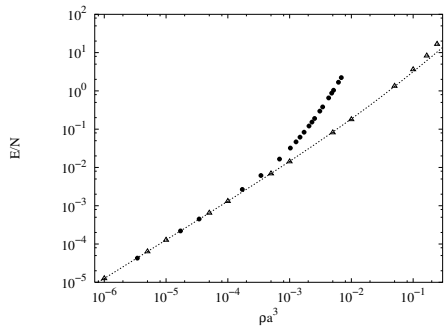}%
        \caption{Equation of state of gas H$\downarrow$ as a function of
	the gas parameter $\rho a^3$ in units of $\hbar^2/(2ma^2)$ (solid
	points). The triangles correspond to a HS gas \protect\cite{giorgini}
	and the line to Eq.(\ref{eq:SFD}).} 
	\label{fig:elowdens}
\end{figure}

The gas phase of spin-polarized H at very low densities is of special
relevance for the field of Bose-condensed gases at low temperatures. The
case of H is even more appealing than alkali gases from a theoretical
perspective because the interatomic interaction between H$\downarrow$ atoms is
very well known, as we have commented in the previous subsection. At
sufficiently low densities, the equation of state becomes universal when
it is written in terms of the gas parameter $x=\rho a^3$, with $a$ the s-wave
scattering length. The equation of state
of a bosonic gas at low densities is given by
\begin{equation}
\left(\frac{E}{N}\right)=4\pi  x \left( 1+ \frac{128}{15 \sqrt{\pi}}
x^{1/2}\right) \ ,
\label{eq:SFD}
\end{equation}
where the first term is the mean-field result,~\cite{universal1} and the
second is the Lee-Huang-Yang correction;~\cite{universal2} the energy per
particle is written in  units of $\hbar^2/(2ma^2)$.

In Fig. \ref{fig:elowdens}, the energy per particle of gas H$\downarrow$ 
is compared to the universal equation of phase (\ref{eq:SFD}) and to DMC 
results 
for a hard-sphere (HS) gas from Ref. \onlinecite{giorgini}. In order to carry out this
comparison we have calculated the s-wave scattering length of the JDW
potential used in the present work. The value obtained, $a=0.697$ \AA\, 
agrees with previous determinations.~\cite{blume,scatlength} As one can see in
the figure, the equation of state of gas H$\downarrow$ coincides with both
the analytic law (\ref{eq:SFD}) and the HS results up to $x \simeq 10^{-4}$
in agreement with the range of universality determined in Ref.
\onlinecite{giorgini}. Beyond this value, the energies of bulk H$\downarrow$
clearly separate from Eq. (\ref{eq:SFD}), increasing with $x$ faster than
for HS gas.

\begin{figure}
\centering
        \includegraphics[width=0.45\linewidth,angle=-90]{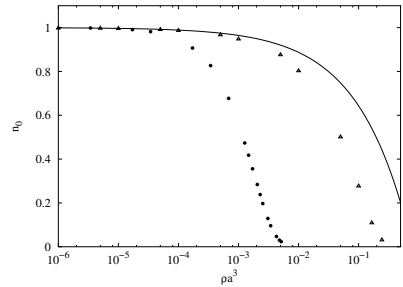}%
        \caption{Condensate fraction of gas H$\downarrow$ as a function of
	the gas parameter $\rho a^3$ (solid
	points). The triangles correspond to a HS gas \protect\cite{giorgini}
	and the line to the Bogoliubov approximation (\ref{bogoliu}).} 
	\label{fig:n0lowdens}
\end{figure}

The condensate fraction $n_0$, i.e., the fraction of particles occupying
the zero momentum state, presents also a universal behavior in terms of the
gas parameter $x$ at very low densities. According to the Bogoliubov
formula,~\cite{universal1}
\begin{equation}
n_0 = 1 - \frac{8}{3 \sqrt{\pi}} x^{1/2}  \ .
\label{bogoliu}
\end{equation}
The condensate fraction has been obtained from the long-range behavior of
the one-body density matrix $\rho(r)$, $n_0=\lim_{r \rightarrow \infty}
\rho(r)$. We have verified by increasing the number of particles of the
simulation at different densities that the size dependence of $n_0$ is
smaller than its statistical error.
In Fig. \ref{fig:n0lowdens}, we compare our results for the condensate fraction
of gas H$\downarrow$ with Eq. (\ref{bogoliu}). DMC results~\cite{giorgini} 
for $n_0$ in the HS system are also plotted in the same figure. As one can
see, the three results are coincident up to  $x \simeq 10^{-4}$, the same
value observed for the energy in Fig. \ref{fig:elowdens}. Both HS and
H$\downarrow$ show a faster decrease with $x$ than the law (\ref{bogoliu}),
the departure from it being significantly larger for hydrogen, in agreement
with the same feature observed in Fig. \ref{fig:elowdens} for the energy.

Spin-polarized H in its gas phase has been studied in all the density
range up to densities above crystallization. DMC results for the total and kinetic energy per 
particle at different densities are reported in Table \ref{tab:Gas}. In
order to remove any residual bias from the trial wave function, kinetic
energies are calculated as differences between  total energies and pure
estimations of potential energies. The energies are positive everywhere,
proving its gaseous nature, and dominated by the kinetic part which is
larger than the potential energy (in absolute values) at any density.   
The potential energy per particle is negative up to a density $\rho=0.015$
\AA$^{-3}$, presenting a minimum value
of -9 K at a density $\rho=0.01$ \AA$^{-3}$, and then becomes positive.

\begin{table}
\begin{center}
\begin{ruledtabular} 
\begin{tabular}{lrrrr}         $\rho$ (\AA $^{-3}$) & 
~$E/N$ (K) & ~~$T/N$ (K) & ~~$P$ (bar) & $c$ (m/s)  \\ \hline 
0.0001 &  0.0221(5)  & 0.101(1) &  0.000321(3) & 19.9(2) \\
0.001  & 0.302(3) & 1.213(6) & 0.0548(4) & 90.8(6) \\
0.005 & 4.091(8) & 9.55(3) & 5.65(5) & 470(4) \\
0.01  & 18.68(6) & 27.40(13) & 61.3(6) & 1149(13) \\
0.0125 & 32.13(6) & 38.61(2)  & 138.4(1.6) & 1.56(2)$\times$10$^3$ \\ 
0.015  & 51.07(8) & 51.9(3) & 273(3) & 2.02(3)$\times$10$^3$ \\
0.02 & 109.24(16) & 83.2(5) & 819(12) & 3.08(5)$\times$10$^3$
\end{tabular} 
\end{ruledtabular}
\end{center}
\caption{Results for gas 
H$\downarrow$ at different densities $\rho$: energy per particle ($E/N$), 
kinetic energy per particle ($T/N$), pressure ($P$), and speed of sound 
($c$). Figures in parenthesis are the statistical errors.} 
\label{tab:Gas}
\end{table}

\begin{figure}
\centering
        \includegraphics[width=0.65\linewidth]{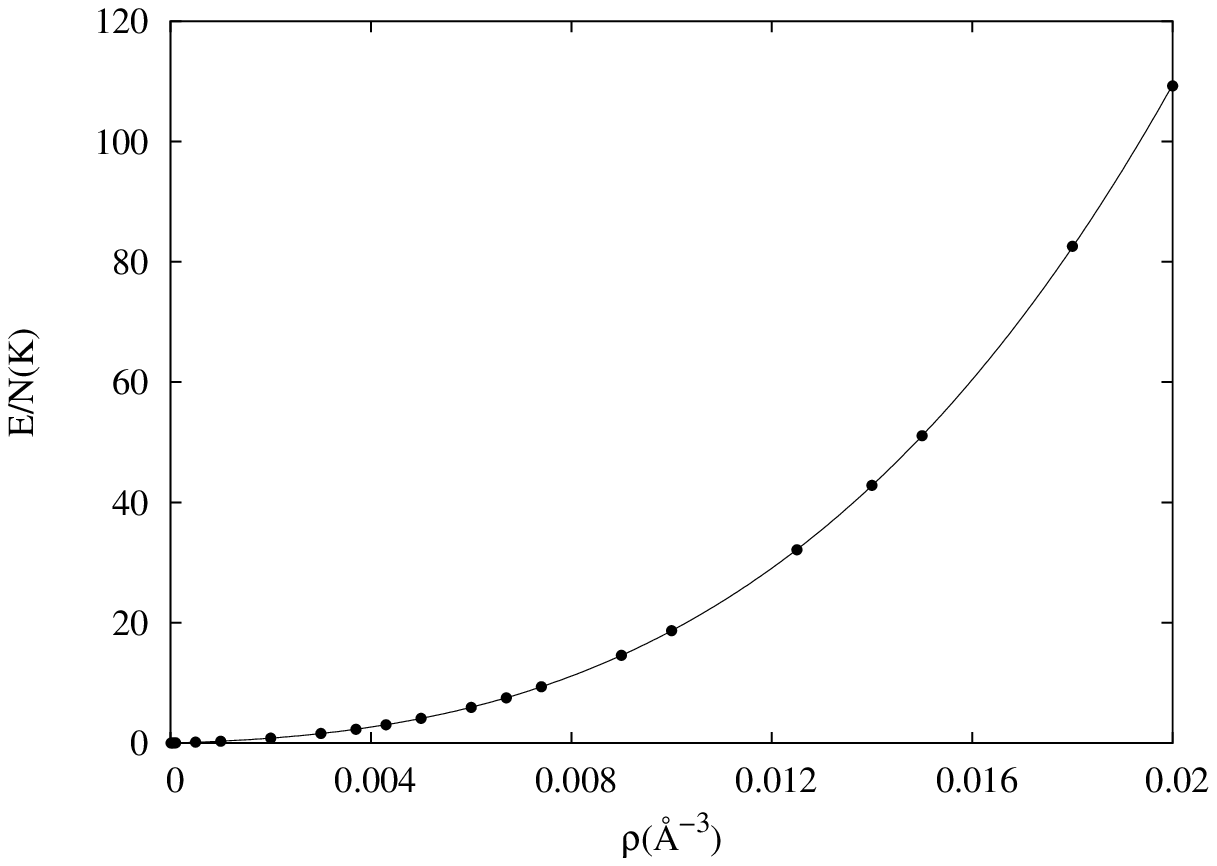}
        \caption{Energy per particle of gas H$\downarrow$ (solid circles) as a
	function of the density $\rho$.  The solid line corresponds to the 
	fit to the DMC energies using 	Eq. (\protect\ref{eqestat}).   
	The error bars of the DMC energies are smaller than the size of the symbols.  }
\label{fig:enliq}	
\end{figure}

In Fig. \ref{fig:enliq}, we plot the
present DMC results for the equation of state of the gas. Our results are
well parameterized by a polynomial form ($e \equiv E/N$)
\begin{equation}
e(\rho)  =  e_1 \rho + e_2 \rho^{2} + e_3 \rho^{3}+e_4 \rho^{4} \ ,
\label{eqestat}
\end{equation}
shown as a solid line on top of the DMC results in Fig. \ref{fig:enliq}.
The best set of parameters is: $e_1=217.0(1.9)$ K\AA, $e_2=7.76(9)\times
10^{4}$ K\AA$^{2}$,  $e_3=8.23(12)\times 10^{6}$ K\AA$^{3}$, and 
$e_3= 5.1(5)\times 10^{7}$ K\AA$^{4}$, the figures in parenthesis being the
statistical uncertainties.

\begin{figure}
\centering
        \includegraphics[width=0.45\linewidth,angle=-90]{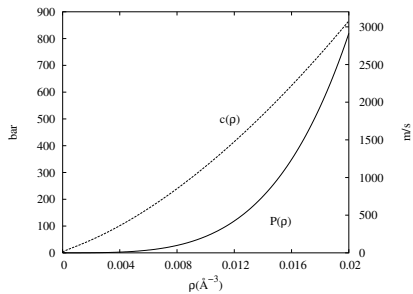}%
        \caption{Pressure and speed of sound of gas H$\downarrow$ 
	as a function of the density. Left (right) scale corresponds to
	pressure (speed of sound).} 
	\label{fig:presgas}
\end{figure}

Using the equation of state (\ref{eqestat}), the pressure is easily derived
from its thermodynamic definition 
\begin{equation}
P(\rho)=\rho^2(\partial e/ \partial \rho) \ ;
\label{pressure}
\end{equation}
and from it, the corresponding speed of sound as a function of the density
\begin{equation} 
c^2(\rho)=\frac{1}{m} \left( \frac{\partial P}{\partial \rho} \right) \ .
\label{speed}
\end{equation} 
Results for the pressure $P$ and the speed of sound $c$ for some values of the
density, where specific DMC simulations have been carried out, are reported in
Table \ref{tab:Gas}. The functions $P(\rho)$ and $c(\rho)$, derived
respectively from Eqs. (\ref{pressure}) and (\ref{speed}), are shown in Fig.
\ref{fig:presgas}.

\begin{figure}
\centering
        \includegraphics[width=0.7\linewidth]{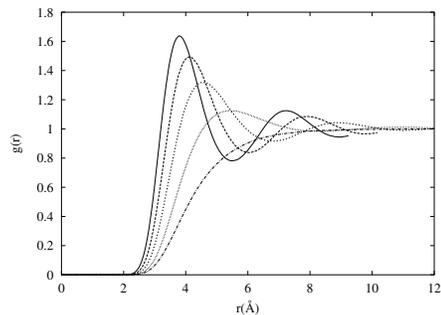}%
        \caption{Two-body radial distribution functions of the gas phase. 
	From bottom to top in the height of the main
	peak, the results correspond to densities 0.002, 0.0067, 0.01,
	0.0125, and 0.015 \AA$^{-3}$.  }
	\label{fig:grfluid}
\end{figure}

DMC simulations allow also for exact estimations of other relevant
magnitudes such as the two-body radial distribution function $g(r)$ and its
Fourier transform, the static structure function $S(k)$. With the use of
pure estimators~\cite{pures} it is possible to eliminate the bias coming from the
trial wave function and arrive to exact results for both functions.
The evolution of $g(r)$ with density for the gas H$\downarrow$ is shown 
in Fig. \ref{fig:grfluid}. At the smallest density reported, $g(r)$ is a
monotonic function with the corresponding
\textit{hole} consequence of the repulsive core of the interatomic
interaction.  When $\rho$ increases $g(r)$ gains structure, with the main peak
that shifts to shorter distances and increases its strength.   

\begin{figure}
\centering
        \includegraphics[width=0.7\linewidth]{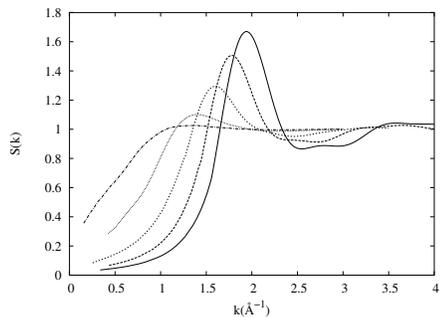}%
        \caption{Static structure function of the gas phase. 
	From bottom to top in the height of the main
	peak, the results correspond to densities 0.002, 0.0067, 0.01,
	0.0125, and 0.015 \AA$^{-3}$.}
	\label{fig:sofk}
\end{figure}

In Fig. \ref{fig:sofk}, results of $S(k)$ at the same densities as in
Fig. \ref{fig:grfluid} are reported.  The results show the expected behavior: 
when $\rho$ increases, 
the strength of the main peak increases and moves to higher momenta  
in a monotonic way. At low momenta, the slope of $S(k)$ decreases with the
density, following the limiting behavior $\lim_{k \rightarrow 0} S(k)= \hbar
k/(2mc)$ driven by the speed of sound $c$. As expected, the DMC data start at a 
finite $k$ value inversely proportional to the box size $L$.    

\begin{figure}
\centering
        \includegraphics[width=0.7\linewidth]{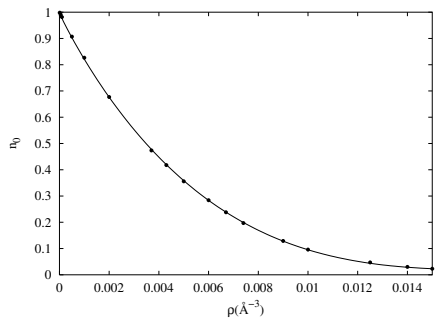}%
        \caption{Condensate fraction of spin-polarized H in the gas phase. 
	The line corresponds to a fit to the DMC data using
	Eq.(\protect\ref{nofit}). The
	error bars are smaller than the size of the symbols.}
	 \label{fig:nogas}
\end{figure}

To end this subsection, we show in Fig. \ref{fig:nogas} the density
dependence of the condensate fraction from the very dilute regime up to
densities corresponding to freezing. The full set of data is well
reproduced using the functional form
\begin{equation}
n_0(\rho) =   1 - \frac{8}{3 \sqrt{\pi}} (\rho a^3)^{1/2}-b_1 \rho a^3-b_2
(\rho a^3)^{2}-b_3 (\rho a^3)^{5/2}  \ ,
\label{nofit}
\end{equation} 
which is also plotted in the figure as a solid line on top of the DMC
results. The values of the parameters in Eq. (\ref{nofit}) are
$b_1=504(5)$,  $b_2=-1.254(49)\times 10^5$, $b_3=8.54(55)\times 10^5$, and $a$ is the scattering length.

\subsection{Solid phase}

The solid phase of spin-polarized H has been studied by using as importance
sampling trial wave function a Nosanow-Jastrow model (\ref{nosanow}). The
geometry of the lattice is defined by a proper selection of the lattice
sites $\bm{r}_I$ around which the atoms are organized according to a
commensurate solid.  There is no experimental measurement on solid
H$\downarrow$ at the pressures in which we are interested in and nothing is
known about the form of its solid lattice at low temperatures. We have
carried out calculations of the solid phase at some densities using the
fcc, hcp, and bcc lattices. Near the melting density the bcc phase is slightly
better and, at higher densities, the differences between them
are not distinguishable within the statistical noise: at $\rho=0.0125$
\AA$^{-3}$, $E/N = 33.12(4)$, $34.30(5)$, and $33.02(8)$ K, and at 
$\rho=0.018$ \AA$^{-3}$, $E/N =76.2(2)$,
$76.2(2)$, and $76.1(2)$ K for fcc, hcp, and bcc respectively. Therefore, we
decided to study the solid H$\downarrow$ properties assuming a bcc phase.
It is worth noticing that the same lattice was used in the past 
by Pierleoni \textit{et al.}~\cite{ceperley} in the
study of solid H at very high pressure.

\begin{table}
\begin{center}
\begin{ruledtabular}                 
\begin{tabular}{lrrrr}
   $\rho$ (\AA $^{-3}$) & ~~$E/N$ (K) 
& ~~$T/N$ (K) & ~~$P$ (bar) & $c$ (m/s)  \\ \hline
0.0125 & 33.02(8) & 43.4(1)  & 121.6(2) & 1445(3) \\
0.015  & 49.38(8) & 59.0(6)      & 236.4(5) & 1866(4) \\
0.02 & 99.6(2) & 93.1(6) & 679.8(1.4) & 2821(6) \\
0.0225 & 134.9(2) & 107.8(6) & 1096(2) & 3350(7) \\         
0.025 & 178.5(2) & 123.7(7) & 1648(3) & 3909(8)                 
 \end{tabular}
  \end{ruledtabular}
  \end{center}         
 \caption{Results for solid H$\downarrow$ at different densities $\rho$: 
energy per particle ($E/N$), kinetic energy per particle ($T/N$), pressure 
($P$), and speed of sound ($c$). Figures in parenthesis are the 
statistical errors.}
\label{tab:BCCHydrogen}
\end{table}

\begin{figure}
\centering
        \includegraphics[width=0.5\linewidth,angle=-90]{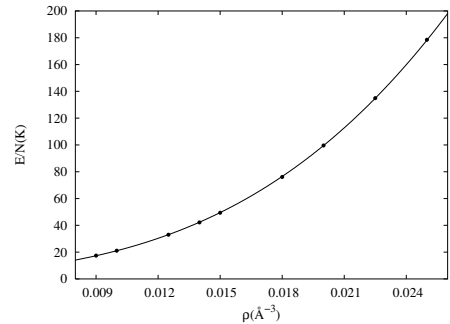}
        \caption{Energy per particle of solid H$\downarrow$ (solid circles) as a
	function of the density $\rho$.  The solid line corresponds to the 
	fit to the DMC energies using 	Eq. (\protect\ref{eqestatsolid}).   
	The error bars of the DMC energies are smaller than the size of the symbols.  }
\label{fig:ensol}	
\end{figure}

Some selected results for the total and kinetic energies per particle are
reported in Table \ref{tab:BCCHydrogen}. The behavior of the partial
energies, potential and kinetic, is very similar to the one obtained for
the gas: the kinetic energy dominates in all the density regime and the
potential energy is negative at the lower densities and becomes positive
for densities $\rho \geq 0.02$ \AA$^{-3}$. The full set of results for the
energy of the solid phase is shown in Fig. \ref{fig:ensol}. The solid line
on top of the DMC results correspond to a numerical fit obtained using the
function
\begin{equation}
e(\rho) =  s_1 \rho +  s_3\rho^{3} \ .
\label{eqestatsolid}
\end{equation}
The optimal values in Eq. (\ref{eqestatsolid}) are $s_1=1147(6)$ K\AA\, 
and $s_3=9.57(2)\times 10^{6}$K\AA$^{3}$.

\begin{figure}
\centering
        \includegraphics[width=0.5\linewidth,angle=-90]{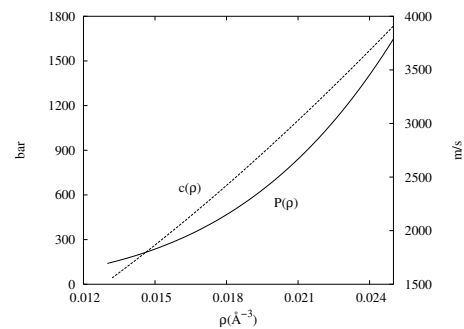}%
        \caption{Pressure and speed of sound of solid H$\downarrow$ 
	as a function of the density. Left (right) scale corresponds to
	pressure (speed of sound).} 
	\label{fig:pressolid}
\end{figure}

From the functional form (\ref{eqestatsolid}), and using the corresponding
thermodynamic expressions for the pressure (\ref{pressure}) and the speed
of sound (\ref{speed}), one can easily derive the dependence of these magnitudes on the density.
The results for both functions are plotted in Fig.
\ref{fig:pressolid}.

\begin{figure}
\centering
        \includegraphics[width=0.5\linewidth,angle=-90]{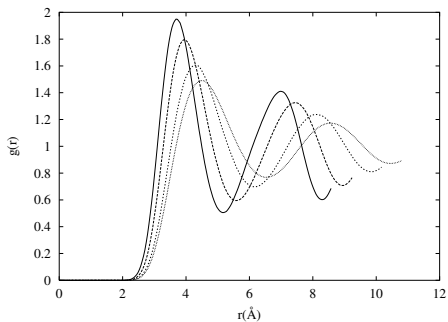}%
        \caption{Two-body radial distribution functions of the solid phase. 
	From bottom to top in the height of the main
	peak, the results correspond to densities 0.0125, 0.015, 0.02,
	and 0.025 \AA$^{-3}$.  }
	\label{fig:grsolid}
\end{figure}

The spatial order of the solid is reflected in the shape of the two-body
radial distribution function $g(r)$. Results for $g(r)$ in the solid phase
for different densities are shown in Fig. \ref{fig:grsolid}. At densities
$\rho=0.0125$ and 0.015 \AA$^{-3}$, we can compare results for $g(r)$ in
the gas and solid phases. As one can see, the peaks of the solid are
slightly shifted to larger distances than in the gas and more importantly,
and as expected, the secondary peaks of the solid are more pronounced. When
$\rho$ increases, the height of the peaks increases and moves to shorter
distances, like in the gas phase.

\subsection{Gas-solid phase transition}
 
\begin{figure}
\centering
        \includegraphics[width=0.7\linewidth]{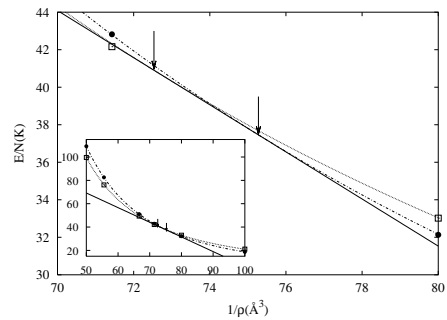}%
        \caption{Maxwell construction based on plotting the energy per
particle, $E/N$ as a function of $1/\rho$. The densities at which the first-order 
transition occurs are identified by finding the common tangent (solid line)
to both the
solid (dotted line) and gas curve (dot-dashed line). The inset shows the
construction in a wider range of $1/\rho$.}
\label{fig:Max}
\end{figure}

A relevant prediction that can be drawn from the present DMC results on the
energies of the gas and solid phases of H$\downarrow$ is the location of
the gas-solid phase transition. 
In order to determine the transition point and the corresponding freezing
($\rho_{\rm f}$) and melting ($\rho_{\rm m}$) densities, we have performed the Maxwell 
double-tangent construction as shown in Fig. \ref{fig:Max}. 
From the common tangent to both phases we obtain   
$\rho_{\rm f}=0.01328$ \AA$^{-3}$ and $\rho_{\rm m}=0.01379$ \AA$^{-3}$, 
which corresponds to a common pressure at the transition of $P=173(15)$
bar. The melting pressure has proven to be quite independent of the lattice
used in the simulation since using fcc and hcp we obtain 175 and 176 bar,
respectively. 
 
At the transition the kinetic energy per particle of the system is
discontinuous. The size of this discontinuity in other quantum fluids  
such as He and Ne
has received the interest in the past from both the theoretical and
experimental sides.~\cite{bafile,simmons,zoppi} In the present system 
we are able to accurately measure this
discontinuity: in the gas side at freezing $T/N=44.0(5)$ K and
in the solid side at melting $T/N=51.5(6)$ K. Therefore, the discontinuity
amounts to 7.5 K approximately. On the other hand, when the system
crystallizes the condensate fraction of the gas is small but not zero,
$n_0=0.04$.

The  Lindemann's ratio, defined as $\gamma=\sqrt{\langle
(\bm{r}-\bm{r}_I)^2\rangle} /a_L$,
where $a_L$ is the lattice constant, can also be obtained from the DMC
simulations. At the melting point it is $\gamma=0.25$, a nearly
identical value to the one of solid $^4$He ($\gamma=0.26$).

\begin{figure}
\centering
        \includegraphics[width=0.5\linewidth,angle=-90]{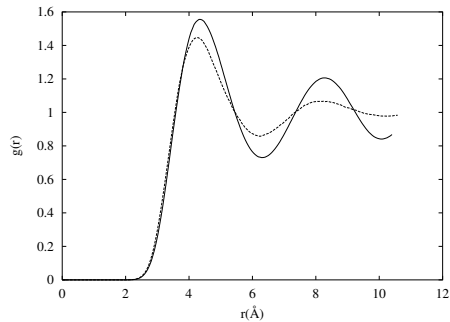}%
        \caption{Two-body radial distribution function at the gas-solid
	phase transition. The solid line corresponds to the solid at
	$\rho_{\rm m}$ and the dashed line to the gas at $\rho_{\rm f}$.}
\label{fig:grtrans}
\end{figure}

\begin{figure}
\centering
        \includegraphics[width=0.5\linewidth,angle=-90]{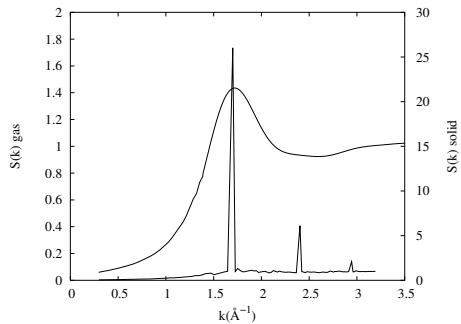}%
        \caption{Static structure factor at the gas-solid
	phase transition. The results correspond to the gas at
	$\rho_{\rm f}$ and to the solid at $\rho_{\rm m}$.}
\label{fig:sktrans}
\end{figure}

The spatial structure of both phases at the transition point is rather
different in spite of the small difference between $\rho_{\rm f}$ and
$\rho_{\rm m}$. In Fig. \ref{fig:grtrans}, we show results of $g(r)$ for
both phases at the transition point. As one can see, the degree of localization is
higher for the solid: the strength of the main peak is larger and the
height of the subsequent peaks decreases more slowly than in the gas.
Nevertheless, the signature of the solid phase manifests more clearly in
$S(k)$. In Fig. \ref{fig:sktrans}, results of $S(k)$ for both phases are
compared at densities  $\rho_{\rm f}$ (gas) and $\rho_{\rm m}$ (solid).
High intensity peaks located at the reciprocal lattice sites are a clear
signature of the solid order; they are obviously absent in the $S(k)$ of
the gas.

\section{conclusions}

The ground-state properties of spin-polarized hydrogen H$\downarrow$ have
been accurately determined using the DMC method. The combination of
the accuracy provided by the DMC and the precise knowledge of the
H$\downarrow$-H$\downarrow$ interatomic potential has allowed for a
nearly exact determination of the main properties of the system, in
both the gas and solid phases. 

The light mass of the hydrogen atom and the shallow well of its 
interaction force H$\downarrow$ to remain a gas in the limit of zero
temperature. In the very dilute limit, the equation of state of the gas is
well described by the general expression of a weakly interacting Bose gas.
This analytical behavior depends only on the gas parameter $\rho a^3$, with
$a$ the corresponding s-wave scattering length. Compared with other
Bose-condensed gases, like the alkalis, hydrogen presents the appealing
circumstance of the accurate knowledge we have of its interatomic
interaction. This allows for the use of the real interaction in all the
density regime, and thus accurate calculations at much higher densities 
are possible. 

When the density is high enough the systems freezes. We have studied the
energetic and structural properties of the solid phase. Near melting the
bcc phase is slightly preferred over the hcp and fcc ones. However, the
energy differences between the  lattices are very small and become
indistinguishable at higher densities. From the DMC equations of state of
the gas and solid phases, we have obtained the gas-solid transition point.
At zero temperature, the transition occurs at $P=173(15)$ bar. This value is
significantly higher than previous estimations: 50 bar, obtained by using
the quantum theory of corresponding states,~\cite{StwaleyNosanow} and 81 bar, 
from a VMC
estimation by Danilowicz \textit{et al.}.~\cite{danilowicz} It is worth
noticing that the transition point depends dramatically on the accuracy of
the theoretical method used for its calculation: if instead of using the
DMC technique an estimation is performed using only 
the VMC method, one obtains $P=113(17)$ bar, a value significantly smaller than the
DMC result.

\acknowledgments
J. B. and J. C. acknowledge support from DGI (Spain) Grant No.
FIS2005-04181 and Generalitat de Catalunya Grant No. 2005SGR-00779.
We also acknowledge the support of the Central Computing Services at
the Johannes Kepler University in Linz, where part of the computations was
performed.

\end{document}